\documentstyle[aabib, epsfig]{l-aa-ps}
\begin{document}
\newcommand{\rf}{\sf}
\newcommand{\E}{{\em Einstein}}
\newcommand{\EMSS}{Extended Medium Sensitivity Survey}
\newcommand{\ESS} {{\em Einstein} Slew Survey}
\newcommand{\EX}{{\em EXOSAT }}
\newcommand{\Mv}{\mbox{$ M_{\rm V} $}}
\newcommand{\mv}{\mbox{$ m_{\rm V} $}}
\newcommand{\bv}{\mbox{$ B-V $}}
\newcommand{\ri}{\mbox{$ R-I $}}
\newcommand{\by}{\mbox{$ b-y $}}
\newcommand{\MH}{\mbox{$[{\rm M}/{\rm H}]$}}
\newcommand{\fxfv}{\mbox{$f_{\rm X}/f_{\rm v}$}}
\newcommand{\fv}{\mbox{$f_{\rm v}$}}
\newcommand{\lx}{\mbox{$L_{\rm X}$}}
\newcommand{\wli}{\mbox{$ W({\rm Li}) $}}
\newcommand{\nli}{\mbox{$ N({\rm Li}) $}}
\newcommand{\nh}{\mbox{$ N({\rm H}) $}}
\newcommand{\cq}{\mbox{$ \chi^2$}}
\newcommand{\XLF}{X-ray luminosity function}
\newcommand{\XLFs}{X-ray luminosity functions}
\newcommand{\vsini}{\mbox{$v \sin(i)$}}
\newcommand{\teff}{\mbox{$T_{\rm eff}$}}
\newcommand{\lii}{\mbox{Li\,{\sc i}}}
\newcommand{\fei}{\mbox{Fe\,{\sc i}}}
\newcommand{\feii}{\mbox{Fe\,{\sc ii}}}
\newcommand{\ali}{\mbox{Al\,{\sc i}}}
\newcommand{\cai}{\mbox{Ca\,{\sc i}}}
\newcommand{\cri}{\mbox{Cr\,{\sc i}}}
\newcommand{\feh}{\mbox{[Fe/H]}}
\newcommand{\lgf}{\mbox{$\log gf$}}
\newcommand{\logg}{\mbox{$\log(g)$}}
\newcommand{\dof}{\mbox{degrees of freedom}}
\newcommand{\muno}{\mbox{$m_1$}}
\newcommand{\dmuno}{\mbox{$\delta m_1$}}
\def\wl{A\hspace{-1.2ex}\raise0.3ex\hbox{\char'27}\hspace{0.5ex}} 

\thesaurus{6(08.06.2; 08.12.1; 08.16.5; 13.25.5)}

\title{On the widespread Weak-Line T-Tauri population detected in the
  ROSAT All-Sky Survey\thanks{Based on observations collected at the
    ESO La Silla observatory}}

\author{F. Favata\inst{1} 
 \and G. Micela\inst{2} 
 \and S. Sciortino\inst{2}}

\institute{Astrophysics Division -- ESA/ESTEC,
  Postbus 299, 2200 AG Noordwijk, The Netherlands 
  \and
  Istituto e Osservatorio Astronomico di Palermo, 
  Piazza del Parlamento 1, 90134 Palermo, Italy 
}

\offprints{F. Favata (ffavata@astro.estec.esa.nl)}

\date{Received 14 January 1997; accepted 15 May 1997}

\maketitle 

\begin{abstract}
    
  We discuss the apparent widespread presence of Weak-Line T-Tauri
  stars (WTTS) among stellar coronal sources detected in the ROSAT
  All-Sky Survey (RASS), and their relative number with respect to
  young main-sequence stars in the same samples. The approach taken in
  most of the current literature for identifying and classifying WTT
  stars among RASS X-ray sources is based on the usage of
  low-resolution optical spectra only and on simple, mass-independent
  thresholds on the equivalent width of the \lii\ 6707.8\,\AA\ 
  doublet. We show that this approach is likely to lead to putative
  WTTS samples which contain a large number of normal, young
  main-sequence stars masquerading as WTTS sources. Young
  main-sequence stars are known to be the dominant contributor in
  stellar X-ray selected samples at the limiting flux levels of the
  RASS, yet they appear to be very rare in the RASS samples discussed
  here. We argue that many of the putative WTTS sources are actually
  mis-classified young main-sequence stars, and that thus there is
  likely not a true ``WTTS question'' in the RASS samples.


\end{abstract}

\section{Introduction}
\label{sec:intro}

X-ray selection has been shown to be a powerful way of selecting
pre-main sequence (PMS) late-type stars, already using \E\ 
observations of star-forming regions (SFR), as for example shown by
\cite*{wvm+94} in the Sco-Cen SFR, where they identified several
previously unknown PMS stars by studying the newly detected X-ray
sources. This technique has recently been extensively applied to the
stellar counterparts of soft X-ray sources detected in the RASS
(\cite{aks+95}, \cite{atw+96}, \cite{akc+97}, \cite{wks+96},
\cite{wkc+97}, \cite{kws+97}, hereafter collectively referred to as
the ``RASS-WTTS papers'') yielding a large number of candidate PMS
stars, which are generally referred to as ``Weak-line T-Tauri Stars''
(WTTS), as they do not show any of the extreme spectral
characteristics (strong emission lines, and large amount of
``veiling'') typical of classical T-Tauri stars (CTTS). The abundance
of PMS stars identified in the RASS, even far away from obvious sites
of star formation, has raised several questions about the mechanisms
of low-mass star formation, about the recent history of star-formation
in the solar neighborhood, and about the mechanisms responsible for
the spatial diffusion of newborn stars from their sites of formation.

These questions have for example been addressed in detail by
\cite*{fei96}, who has discussed various possible models for the
diffusion of young stars from their place of birth, to match the WTTS
population identified around the Chameleon star forming region
investigated by \cite*{aks+95}. One persistent difficulty observed by
\cite*{fei96} is that the putative age of the RASS WTTS population is
consistently too young. One proposed way of accounting for such a
large number of very young stars far away from their plausible
birthplace is to assume the presence of a large number of spatially
sparse, small sites of star formation which have, by the time these
stars have been observed, dissipated away. The concentration of
low-mass star formation in these small sites would challenge much of
our understanding of low-mass star formation. As \cite*{fei96}
remarks, these explanations must be considered as tentative, as the
possibility that a fraction of the RASS high-lithium stars are already
on the main sequence cannot be discarded.

\cite*{bhs+97} reach, for the RASS stars, the same conclusion which
was reached earlier for the \E\ Extended Medium Sensitivity Survey
(EMSS) by \cite*{msf93}, namely that the majority of low-mass stars
detected in X-ray flux-limited surveys at the flux levels typical of
the RASS and of the EMSS are likely to be young main sequence stars,
rather than PMS stars. To support this statement, \cite*{bhs+97} use a
numerical approach, similar to the one presented by \cite*{fms+92},
and applied by \cite*{msf93} and \cite*{sfm95} to the data from the
EMSS, similarly concluding that the majority of the low-mass stars
detected in the RASS are likely to be young main-sequence stars.

Very similar conclusions are reached by \cite*{ghm+96} who have
developed a similar (although fully independent) model of galactic
coronal X-ray source counts. Their work includes a more detailed
modeling of the scale-height evolution of the young stellar
populations, and thus succeeds in better predicting the low-latitude
population. This model has been shown to match well the observations
of low-latitude RASS fields (\cite{mgh+97}), again showing the
predominance of young main-sequence coronal sources.

The large numbers of WTTS identified in the RASS-WTTS papers is in
contrast with the small number of young (from a few times $10^{7}$ up
to $10^8$\,yr of age) main-sequence stars in the same samples. For
example, the low number of non-PMS coronal sources detected by
\cite*{aks+95} contrasts sharply with the much larger number of such
sources detected in the RASS by \cite*{mgh+97}. The contrast is even
more striking if one considers that the latter sample has a shallower
limiting flux. An analysis of the RASS-WTTS papers shows that the
attribution of WTTS ``nature'' has been done on the basis of
low-resolution spectra, and that no lithium abundance determination
has been used to assess the (eventual) PMS status of individual
sources. Either the eventual presence of a ``strong'' \lii\ doublet
(the watershed feature used to discriminate between ``WTTS'' and
``other active stars'') is quoted as determining the ``WTTS nature''
of a source (without any quantitative definition of ``strong'', as in
\cite{aks+95}) or a single mass-independent, minimum equivalent width
of the \lii\ doublet is taken as discriminating between WTTS and other
active stars (as in \cite{wks+96}). In all the above works, the
equivalent width of the \lii\ doublet is measured using spectra with
resolution ranging from $\simeq 4$\,\AA\ to $\simeq 10$\,\AA.

Experience shows that the detection, and, a fortiori, the measurement
of weak spectral features in low-resolution spectra is an uncertain
operation, and that the measurement of spectral features whose
equivalent width is a small fraction of the spectral resolution is
likely to be fraught with large systematic as well as statistical
errors. To assess the reliability of a WTTS identification process
based only on low-resolution spectra, we have analyzed the low-
($\simeq 4$\,\AA) and high-resolution ($\simeq 0.1$\,\AA) spectra of a
number of active stars spanning a wide range of spectral types and
lithium abundances. We have also studied the influence of the usage of
a single watershed value for the equivalent width of the \lii\ feature
as determining whether a source is a WTTS or not, noting the
systematic biases in the resulting samples of X-ray selected WTTS.

The present paper is structured as follow: Sect.~\ref{sec:sample}
briefly presents the sample of stars whose spectra have been studied
here and describes the data reduction, Sect.~\ref{sec:anal} discusses
the derivation of \lii\ doublet equivalent widths and compares in
detail the results obtained from low- and high-resolution spectra,
Sect.~\ref{sec:water} discusses the effects of using a simple
threshold in equivalent width of the \lii\ doublet in the
identification of PMS sources, and Sect.~\ref{sec:young} discusses the
apparent lack of main-sequence stars in the samples from the RASS-WTTS
papers. Finally, Sect.~\ref{sec:disc} discusses the implications of
the findings of the present paper.

\section{The observed sample}
\label{sec:sample}

Our sample consists of a number of late-type stars which we have
previously investigated in the context of our study of X-ray selected
stars from the EMSS and the \E\ Slew Survey (ESS). For all these stars
we had available both low- and high-resolution spectra. The stars were
selected to span a range of spectral types, as well as of lithium
abundance. The sample stars have spectral types ranging from G0 to M0,
and have lithium abundances (as determined on the basis on
high-resolution spectroscopy by \cite*{fbm+95}) ranging from below the
detection limit in high-resolution spectra (implying equivalent widths
of the \lii\ 6707.8\,\AA\ doublet smaller than $\simeq 10$\,m\AA) up
to cosmic lithium abundance ($\nli \simeq 3.2$).  Their spectra are
representative of the larger EMSS and ESS samples.  The
characteristics of the sample stars are listed in
Table~\ref{tab:sample}. In addition, we have studied three M stars
which are listed in Table~\ref{tab:samplem} and are discussed in
detail in Sect.~\ref{sec:m}. This small sample of M stars, includes a
main-sequence, very active star as well as two bona fide PMS stars,
and, while not homogeneous in selection criteria with the first
sample, it supplements it at the cooler end.

\begin{table*}[thbp]
  \caption{The stars for which we have measured the equivalent width
    of the \lii\ 6707.8\,\wl\ doublet in both high-resolution
    (0.05\,\wl\ per pixel) and low-resolution (2\,\wl\ per pixel)
    spectra. The equivalent width of the real \lii\ doublet measured
    in the high-resolution spectra is taken from Favata et al. (1995)
    while the equivalent width of the feature at $\simeq 6708$\,\wl\ 
    visible in all the low-resolution spectra has been measured in the
    present work. The projected rotational velocity of each source
    (which gives a measure of the eventual rotational broadening of
    the lines), the lithium abundance and the \teff\ are also from
    Favata et al. (1995). The last column gives the difference between
    the true and measured (on the low-resolution spectra) distance
    between 6708\,\wl\ \lii\ feature and the 6717\,\wl\ \cai\ feature.
    In all cases this difference is significantly smaller than the
    spectral resolution of $\simeq 4$\,\wl, indicating the the
    identification of the measured feature with the lithium feature at
    6708\,\wl\ appears to be reliable.}
\label{tab:sample}
\begin{flushleft}
\scriptsize
\begin{tabular}{lllrrrrrrr} \hline \\[-0.5pt]
  \multicolumn{1}{c}{\E\ name} & \multicolumn{1}{c}{Other name} &
  \multicolumn{1}{c}{Spectral} & \multicolumn{1}{c}{\teff} 
  & \multicolumn{1}{c}{\wli, \AA} & \multicolumn{1}{c}{\nli} &  
  \multicolumn{1}{c}{\vsini}&
  \multicolumn{1}{c}{\wli. \AA} &  \multicolumn{1}{c}{$\Delta \lambda$} \\
  & & \multicolumn{1}{c}{type} & & \multicolumn{1}{c}{(High res.)} &
   & \multicolumn{1}{c}{km/s} & \multicolumn{1}{c}{(Low res.)} &
  \multicolumn{1}{c}{\AA} \\ \hline
  \\[0.5pt]
1ES0412$+$06.0A & HD26923 & G0V   & 5970 & 0.09  &  2.78       &$<8$&0.32 & $+0.41  $ \\
1ES0637$-$61.4  & HD48189 & G1.5V & 5970 & 0.13  &  3.30       & 15 &0.45 & $-1.49   $ \\
1ES0635$-$69.8 & HD47875 & G3V   & 5720 & 0.20  &  3.70       & 10 &0.46 & $-0.46$ \\
1ES1044$-$49.1 & HD93497  & G5III & 4860 &$\le 0.01$&  $\le -0.30$& -- &0.42 & $-0.39 $ \\
1ES0412$+$06.0B & HD26913 & G8V   & 5540 & 0.06  &  2.14       &$<8$&0.40 & $+0.38 $ \\
1ES0357$-$40.0 & HD25300 & K0e   & 5240 & 0.11  &  2.26       & 12 &0.48 & $+0.33 $ \\
1ES0250$-$12.9 & HD17925 & K1V   & 5050 & 0.20  &  2.88       &$<8$&0.42 & $-0.17 $ \\
1ES0327$-$24.2 & HD21703 & K4V   & 4460 &$\le 0.01$&  $\le 1.74$ &$<8$&0.25 & $-1.08  $ \\
1ES0457$+$01.7A & GJ182 & M1Ve  & 4020 & 0.27  &  1.77       & 14 &0.36 & $-0.43  $ \\
\hline\\ [2pt]
\end{tabular}
\end{flushleft}
\end{table*}

The low-resolution spectra used here have all been acquired using the
ESO 1.5\,m spectroscopic telescope with the Boller \& Chivens
spectrograph. The combination of grating and CCD chip used yielded a
resolution of 1.9\,\AA\ per pixel, or a two-pixel resolution of
$\simeq 4$\,\AA, very similar to the higher resolution employed by
\cite*{aks+95}. The high-resolution spectra were acquired using the
ESO CAT 1.4\,m telescope with the Coud\'e Echelle Spectrometer (CES).
The short camera with the RCA CCD (ESO \#9) was used, yielding an
effective resolution of about 50\,000 (or $\simeq 0.05$\,\AA\ per
pixel). The spectra are centered on the \lii\ 6707.8\,\AA\ doublet,
and cover the range $\simeq 6690$--$6730$\,\AA. The data reduction
procedure has been described in detail in \cite*{fbm+93}, to which the
reader is referred for details.  These spectra allow a measurement of
the equivalent width of the \lii\ 6707.8\,\AA\ doublet to a precision
of a few m\AA\ (depending on the rotational velocity), and thus form
an excellent benchmark for checking the possibility of measuring the
equivalent width of the same line at lower resolutions.

\section{Derivation of lithium equivalent widths from low-resolution
  spectra}
\label{sec:anal}

\begin{figure}[thbp]
  \begin{center}
    \leavevmode
    \epsfig{file=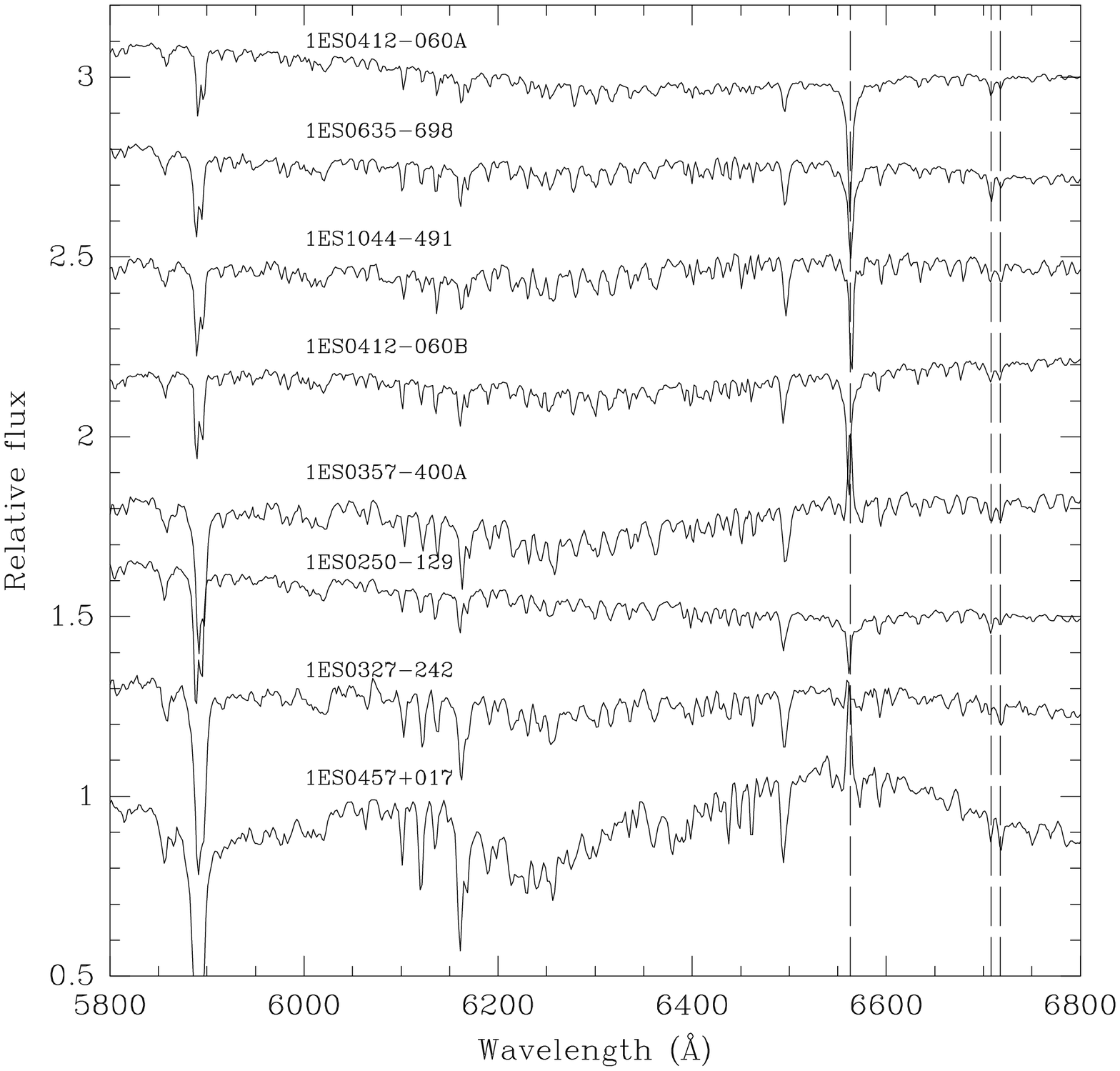, width=8.8cm, bbllx=0pt, bblly=0pt,
      bburx=560pt, bbury=550pt, clip=}
  \end{center}
  \caption{A segment of the low-resolution (2\,\wl\ per pixel) spectra
    for the active stars of Table~\protect{\ref{tab:sample}}. The
    positions of the \lii\ 6707.7\,\wl\ doublet, as well as the
    position of the nearby \cai\ 6717\,\wl\ line are marked by the
    vertical dashed lines. Also marked is the position of the
    H$\alpha$ line. 1ES0637$-$61.4 has not been plotted to avoid
    excessive crowding.}
  \label{fig:lr} 
\end{figure}

\begin{figure}[thbp]
  \begin{center}
    \leavevmode
    \epsfig{file=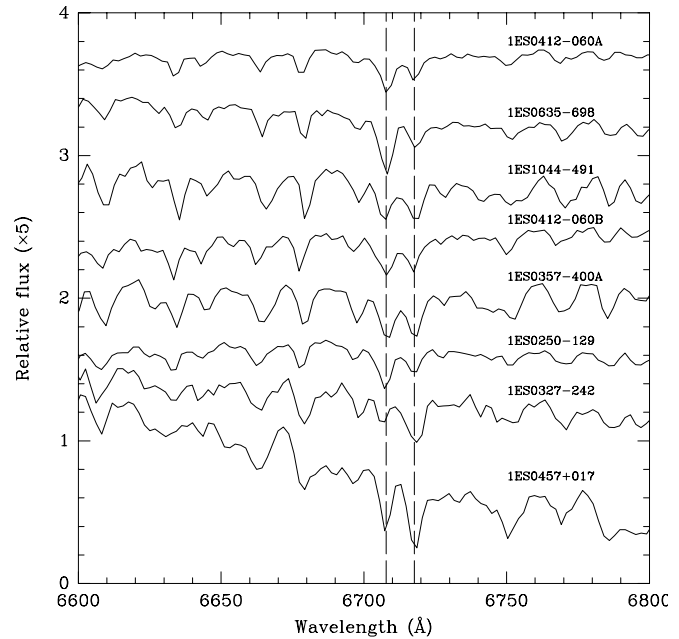, width=8.8cm, bbllx=0pt, bblly=0pt,
      bburx=560pt, bbury=550pt, clip=}
  \end{center}
  \caption{The region of the \lii\ doublet, from the same spectra as
    in Fig.~\protect{\ref{fig:lr}}, vertically expanded by a factor of
    5 to better show the spectral features. The positions of the \lii\ 
    6707.7\,\wl\ doublet, as well as the position of the nearby \cai\ 
    6717\,\wl\ line are again marked by vertical dashed lines.}
  \label{fig:lr2}
\end{figure}

A representative segment of the low-resolution spectra used in the
present work is shown in Fig.~\ref{fig:lr}, in which the expected
positions of the \lii\ 6707.8\,\AA\ doublet and of the nearby \cai\ 
6717.7\,\AA\ line are shown by vertical dashed lines.
Figure~\ref{fig:lr2} shows an enlargement of the same spectra, trimmed
to cover a small region near the \lii\ doublet, and expanded
vertically by a factor of 5, for clarity. All the stars in the sample
appear to have a spectral feature in absorption at the expected
position of the \lii\ doublet.

We have measured the equivalent width of the line present in the low
resolution spectra at the expected position of the \lii\ doublet,
using the IRAF {\sc splot} task, by fitting two gaussians, one to the
line identified with the \lii\ doublet itself, and the other to the
nearby \cai\ feature, which is often blended with the \lii\ feature.
The best-fit equivalent widths are reported in Table~\ref{tab:sample},
together with the difference between the true and measured (on the
low-resolution spectra) distance between 6708\,\AA\ \lii\ feature and
the 6717\,\AA\ \cai\ feature.  This last quantity can be used as a
measure of the reliability of the identification of the feature near
6708\,\AA\ with the \lii\ line, as it should be significantly smaller
than the spectral resolution (it indeed is for all the program stars).

We have found the measurement of such weak features (few hundreds of
m\AA\ at most) on spectra of such low resolution to be an uncertain
process. The lack of any line-free continuum in the neighborhood of
the line in question makes continuum estimation a subjective process,
and we estimate the uncertainty due to placement of the continuum
alone to be at least some 100\,m\AA. Such uncertainty is in line with
the uncertainty quoted by \cite*{aks+95}, for their measurements on
similar spectra, of 150\,m\AA\ (i.e. comparable, or often larger than
the equivalent width being measured. Their quoted uncertainty is
independent of the spectral resolution).

All the equivalent widths measured in the low-resolution spectra for
the feature at $\simeq 6708$\,\AA\ are higher than the true equivalent
width of the \lii\ doublet, sometimes by several hundreds m\AA, and
would, taken at face value, imply (following the criteria of the
RASS-WTTS papers) that all the sources discussed here are PMS, or,
more specifically, WTT stars. Yet, none of these stars shows evidence
of being a WTTS when real lithium abundances are derived from
high-resolution spectra of the same stars. The two later-type
high-lithium sources in our sample, 1ES0250$-$12.9 and
1ES0457$+$01.7A, which could be suspected of being PMS stars, while
certainly young, are also very close to the main-sequence, and
certainly not any longer on the Hayashi track, as unambiguously shown,
on the basis of Hipparcos parallaxes, by \cite*{mfs97}.

\begin{figure}[htbp]
  \begin{center}
    \leavevmode
    \epsfig{file=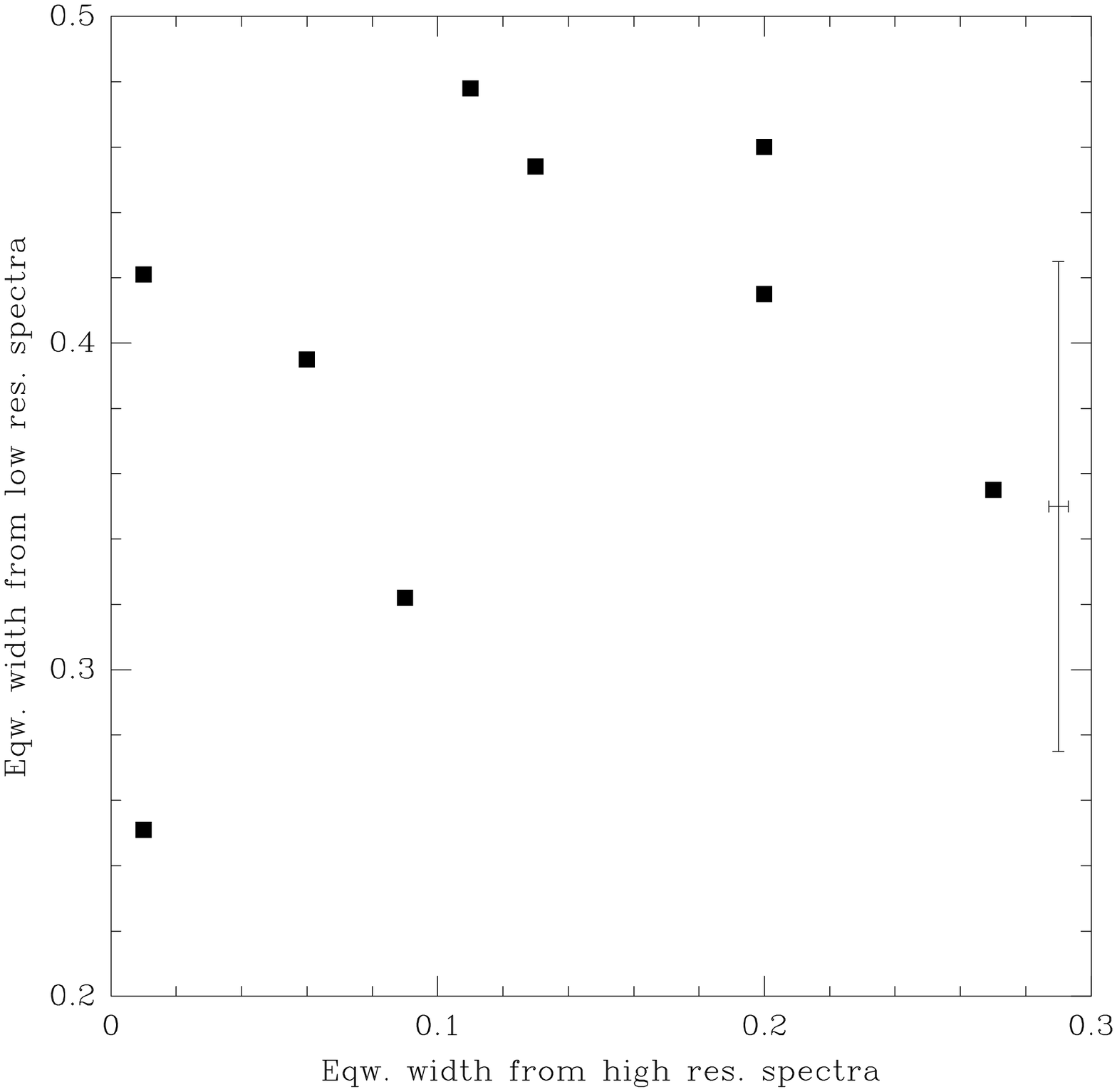, width=8.8cm, bbllx=0pt, bblly=0pt,
      bburx=560pt, bbury=550pt, clip=}
  \end{center}
  \caption{The equivalent width of the 6708\,\wl\ feature derived from
    the low-resolution spectra plotted against the equivalent width of
    the \lii\ doublet from the high-resolution spectra for the program
    stars. Also plotted is a typical error bar (assuming a 150\,m\wl\ 
    uncertainty for the low-resolution measurements and a 5\,m\wl\ 
    uncertainty for the high-resolution measurements).}
  \label{fig:scatt}
\end{figure}

The measurement of the 6708\,\AA\ feature in low-resolution spectra of
low-mass stars is thus very likely to lead to a significant
over-estimate the true lithium abundance. Even worse, two
sources (1ES0327$-$24.2 and 1ES1044$-$49.1) which have no measurable
lithium down to less than 10\,m\AA\ in their high-resolution spectra,
appear to have a similar feature at 6708\,\AA\ as the stars with deep
\lii\ doublets visible in their low-resolution spectra. Furthermore,
there seems to be no clear relationship between the equivalent width
of the 6708\,\AA\ feature measured in low-resolution spectra and the
equivalent width of the \lii\ doublet, as the very large measurement
error makes it impossible to simply subtract the (metallicity and
effective temperature dependent) ``foot'' due to the contribution of
the \fei\ lines to the feature measured in the low-resolution spectrum
deriving the ``true'' equivalent width of the \lii\ doublet. This is
clearly illustrated in Fig.~\ref{fig:scatt}, which shows a scatter
plot of the equivalent width of the the \lii\ feature as measured in
low-resolution spectra as a function of its ``true'' equivalent width.

\begin{figure}[htbp]
  \begin{center}
    \leavevmode
    \epsfig{file=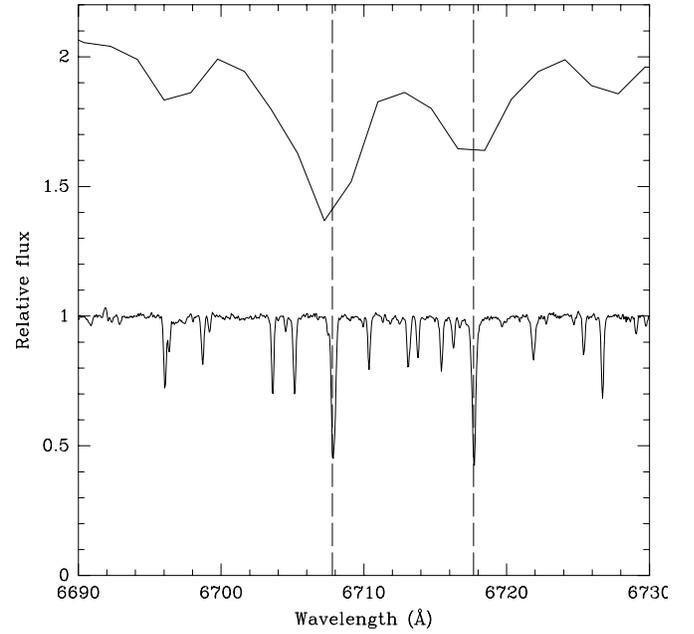, width=8.8cm, bbllx=0pt, bblly=0pt,
      bburx=560pt, bbury=550pt, clip=}
  \end{center}
  \caption{High- and low-resolution spectra of the region around the
    \lii\ 6707.8\,\wl\ doublet for the K1 dwarf 1ES0250$-$12.9. The
    vertical lines mark the position of the \lii\ 6707.8\,\wl\ and
    \cai\ 6717\,\wl\ features.}
  \label{fig:li}
\end{figure}

Figure~\ref{fig:li} shows, superimposed, the high- and low-resolution
spectra of 1ES0250$-$12.9, a high-lithium K1 dwarf. Both spectra are
shifted to the rest wavelength of the spectral lines. As it is obvious
from the plot, both the ``\cai'' and the ``\lii'' feature in the low
resolution spectrum are actually a blend of several spectral features,
making it impossible to accurately derive the true equivalent width of
the parent lines.

\begin{figure}[htbp]
  \begin{center}
    \leavevmode
    \epsfig{file=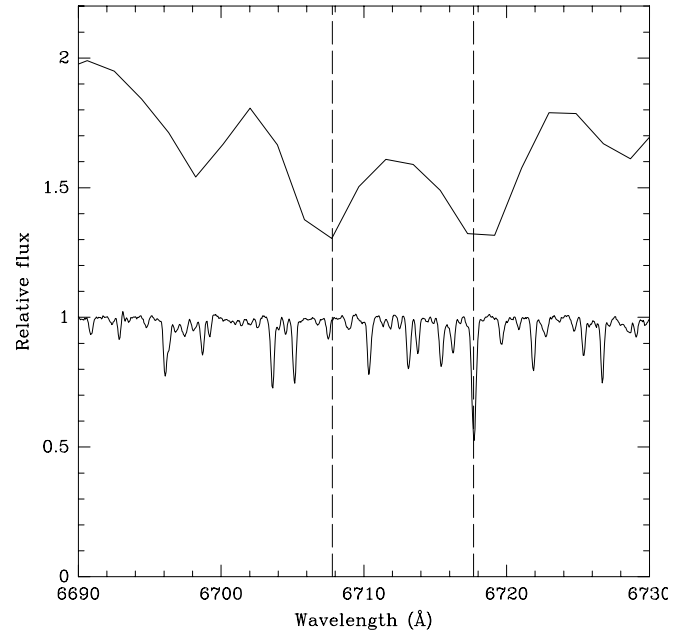, width=8.8cm, bbllx=0pt, bblly=0pt,
      bburx=560pt, bbury=550pt, clip=}
  \end{center}
  \caption{High- and low-resolution spectra of the region around the
    \lii\ 6707.8\,\wl\ doublet for the G5 giant 1ES1044$-$49.1. The
    vertical lines mark the position of the \lii\ 6707.8\,\wl\ and
    \cai\ 6717\,\wl\ features.}
  \label{fig:noli}
\end{figure}

Figure~\ref{fig:noli} shows the same type of spectra for
1ES1044$-$49.1, a G5 giant which has (as evident from the high
resolution spectrum) no measurable lithium. Remarkably, the \fei\ 
lines evident in the high-resolution spectrum ``bunch'' together, in
the low-resolution spectrum, to mimic a feature at a wavelength not
distinguishable (at these resolutions) from a feature containing a
sizable contribution from the \lii\ doublet, and which would thus be
confused with it in the absence of high-resolution spectra.

The strength of the 6708\,\AA\ feature visible in the low-resolution
spectra of G and K stars thus appears not to be a reliable indicator
of the equivalent width of the lithium doublet, and bears little
relationship with the actual lithium abundance of the source.  The
presence of an absorption feature at 6708\,\AA\ in low-resolution
spectra, should thus not, per se, be taken as an indication of the
possible PMS status of a G- or K-type star. Further studies on both
the low- and high-resolution spectra of large samples would be needed
to asses whether, in the presence of a very accurate wavelenght
solution for the low-resolution spectrum, the lithium-mimicking
feature visible in Li-poor stars could be reliably distinguished from
a true \lii\ feature. Even if this were the case, however, the problem
of the associated large uncertainty in the derived equivalent width
(related to the low resolution) would still stand.

\subsection{M stars}
\label{sec:m}

\begin{figure}[htbp]
  \begin{center}
    \leavevmode
    \epsfig{file=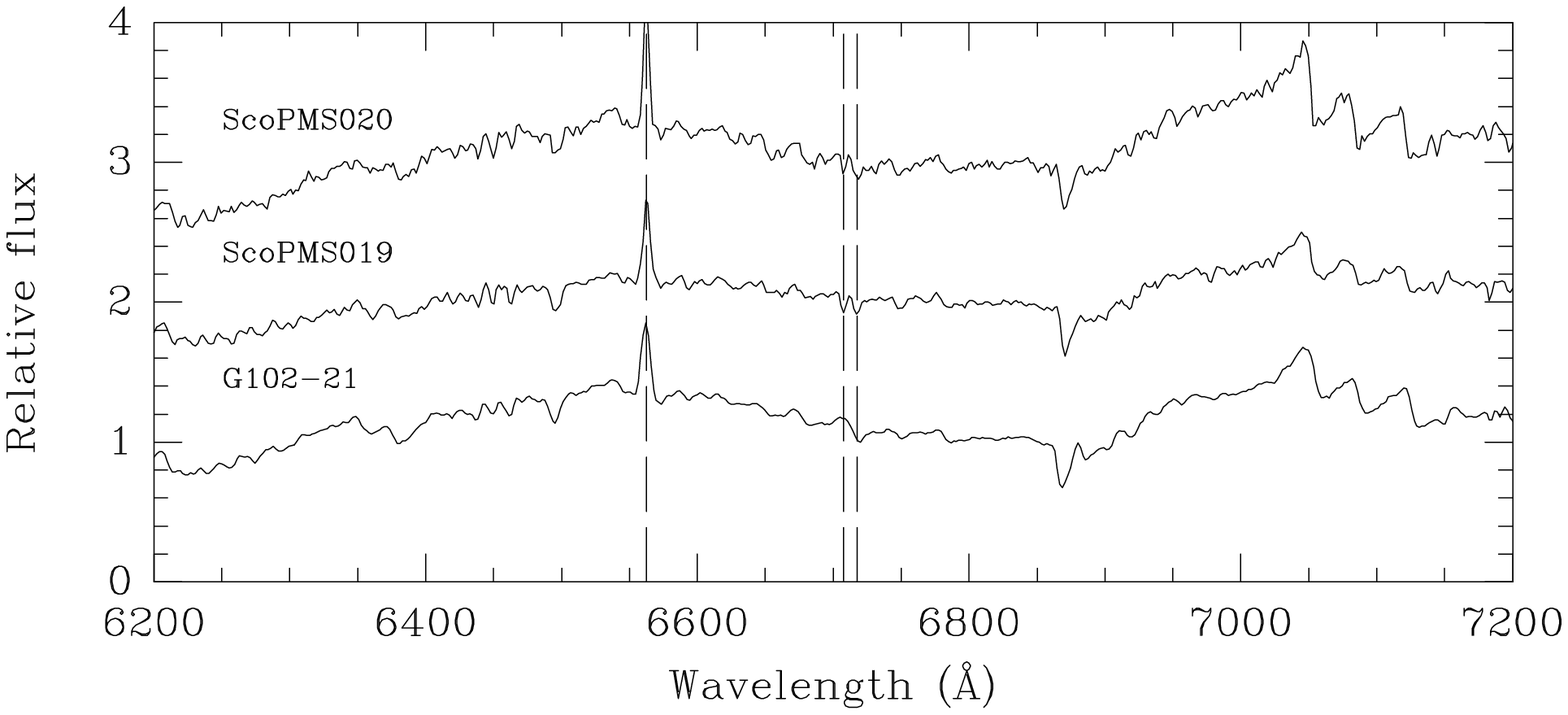, width=8.8cm, bbllx=0pt, bblly=0pt,
      bburx=560pt, bbury=300pt, clip=}
  \end{center}
  \caption{A segment of the low-resolution ($\simeq 2$\,\wl\ per
    pixel) spectra for the three M stars of
    Table~\protect{\ref{tab:samplem}}. The positions of the \lii\ 
    6707.7\,\wl\ doublet, of the nearby \cai\ 6717\,\wl\ line and of
    the H$\alpha$ line are marked by vertical dashed lines.}
  \label{fig:m2}
\end{figure}

\begin{figure}[htbp]
  \begin{center}
    \leavevmode
    \epsfig{file=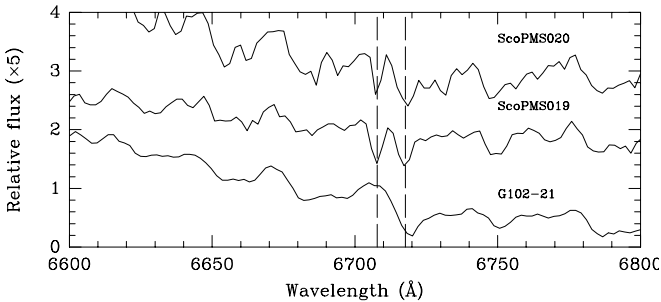, width=8.8cm, bbllx=0pt, bblly=0pt,
      bburx=560pt, bbury=300pt, clip=}
  \end{center}
  \caption{The region of the \lii\ doublet, from the same spectra as
    in Fig.~\protect{\ref{fig:m2}}, vertically expanded by a factor of
    5 to better show the presence of lines. The positions of the \lii\ 
    6707.7\,\wl\ doublet and of the nearby \cai\ 6717\,\wl\ line are
    again marked by vertical dashed lines.}
  \label{fig:m}
\end{figure}

M stars are likely to be easier targets for spotting WTTS sources from
low-resolution spectra. In cooler stars most metallic lines (such as
the \fei\ lines in the region around the \lii\ doublet) become weak,
and merge in a maze of molecular lines (mostly from TiO and from
hydrides, specially MgH), forming, in a low-resolution spectrum, a
pseudo-continuum against which a strong \lii\ doublet is likely to
stand out. At the same time, for a given lithium abundance the
equivalent width of the \lii\ doublet will be stronger, because of the
lower ionization fraction of lithium at lower temperature. While we do
not have available, for M stars, an extensive library of both high-
and low-resolution spectra, we had a few low-resolution spectra of
active M stars with no measurable lithium in their high-resolution
spectra and of bona fide M-type WTTS with a strong lithium line
visible in high-resolution spectra. At the low resolution discussed
here (also $\simeq 4$\,\AA) these stars appear to be easily
distinguishable. The low-resolution spectra of three of these M stars
are shown in Figs.~\ref{fig:m2} and~\ref{fig:m}, which are analogous
to Figs.~\ref{fig:lr} and~\ref{fig:lr2} shown earlier for G and K
stars. The bottom star in both figures (G102$-$21, which has been
discussed by \cite{mfp+95}) has no measurable lithium line in its
high-resolution spectrum (and has no 6708\,\AA\ feature in its
low-resolution spectrum), while the two top spectra are from bona fide
WTT stars in the Sco-Cen SFR reported by \cite*{wvm+94}, and have
clearly visible 6708\,\AA\ features. The presence of strong lithium in
absorption in their spectra has been determined by \cite*{wvm+94} on
the basis of high-resolution spectra, and it is reported in
Table~\ref{tab:samplem}. The low-resolution spectrum of G102$-$21 is
very similar to the spectra of the WTTS, the only visible difference
in their spectra being the \lii\ doublet in absorption.

\begin{table}[tbhp]
  \caption{The three M stars, two bona fide WTTS and an active M star
    with no lithium in its high-resolution spectra, for which spectra
    are shown in Fig.~\protect{\ref{fig:m2}}
    and~\protect{\ref{fig:m}}.  The high-resolution \lii\ equivalent
    widths and rotational velocity are from Walter et al.  (1994) for
    the two WTTS and from our own work (see Micela et al. 1995) for
    G102$-$21. $\Delta \lambda$ is defined as in
    Table~1.}
\label{tab:samplem}
\begin{flushleft}
\scriptsize
\begin{tabular}{llrrrrr} \hline \\[-0.5pt]
  \multicolumn{1}{c}{Name} &
  \multicolumn{1}{c}{Spectral} & 
  \multicolumn{1}{c}{\wli, \AA} &
  \multicolumn{1}{c}{\wli, \AA} & \multicolumn{1}{c}{\vsini} & 
  \multicolumn{1}{c}{$\Delta \lambda$}\\
  & \multicolumn{1}{c}{type} & 
  \multicolumn{1}{c}{(High res.)} &
  \multicolumn{1}{c}{(Low res.)} &\multicolumn{1}{c}{km/s} &
  \multicolumn{1}{c}{\AA} \\ \hline
  \\[0.5pt]
  ScoPMS020    & M3 & 0.49       & 0.55      & 27 & $-0.8$   \\
  ScoPMS019    & M1 & 0.61       & 0.49      & 19 & $+0.5$   \\
  G102-21      & M2 & $\le 0.01$ & $\le 0.1$ & 20 & --   \\
\hline\\ [2pt]
\end{tabular}
\end{flushleft}
\end{table}

The measured equivalent width of the \lii\ doublet in low- and
high-resolution spectra for M stars are compatible within the large
observational error of the low-resolution measurements. For M stars
low-resolution spectra thus appear to be a useful tool to search for
high-lithium stars.  Again, however, any quantitative attempt at
measuring the equivalent width of the \lii\ doublet on low-resolution
spectra is still likely to be affected by large errors, whose amount
will depend both on the spectral resolution used and on the precise
spectral type and metallicity of the star being observed.

\section{Lithium equivalent width, age, and T-Tauri nature}
\label{sec:water}

The attribution of a precise age to individual stars is difficult,
specially when individual accurate distances are not available, which,
specially for young stars, allow the placement on evolutionary tracks.
The only available indicator of WTTS status for individual low-mass
stars of unknown distance is the presence of a very large lithium
abundance: low-mass stars are supposed to burn lithium at the basis of
their convective zones, with lower mass stars having proportionally
deeper convection zones and thus shorter lithium depletion
characteristics times. The available observational evidence, however,
is not so simple, and all the currently available data point toward
several factors, in addition to age, influencing the lithium abundance
of a low-mass star.

Old dwarfs earlier than G5 show a large range of lithium abundance,
with lithium-rich stars being found at essentially any age
(\cite{plp94}, \cite{fms96a}), and the same stars still have
essentially undepleted lithium when they are well in the main sequence
stage (as in the Pleiades, \cite{sjb+93}), showing that the lithium
criterium is of little relevance for their being classified as PMS
sources. For low-mass stars cooler than $\simeq$ G5, the evidence so
far available points toward lithium being depleted with age, although
young stars have a wide range of lithium abundance at any given age,
as shown, for example, by \cite*{sjb+93} for the solar-type stars in
the Pleiades, and by \cite*{spg+93} for the $\alpha$ Persei cluster.

Thus, while high levels of lithium (i.e.\ comparable to the ``cosmic''
abundance, $\nli \simeq 3.2$ on the usual scale where $\nh = 12.0$)
are characteristic of G- and K-type PMS stars (and are a ``necessary''
condition for being classified as a PMS), a lithium abundance of order
$\simeq 3.0$ is by no means sufficient for classifying a star as PMS,
given than many stars down to mid-K spectral type in the Pleiades have
a lithium abundance close to 3.0, yet they clearly are on (or very
close to) the main sequence.

Given that cooler stars are expected (and generally observed) to be
depleting their lithium more rapidly than higher mass solar-type
stars, it is not possible to adopt a single, mass-independent
threshold of the lithium abundance as discriminating between PMS and
main-sequence stars, with the same lithium abundance having quite
different implications on the evolutionary status in G, K or M stars.

\begin{figure}[htbp]
  \begin{center}
    \leavevmode
    \epsfig{file=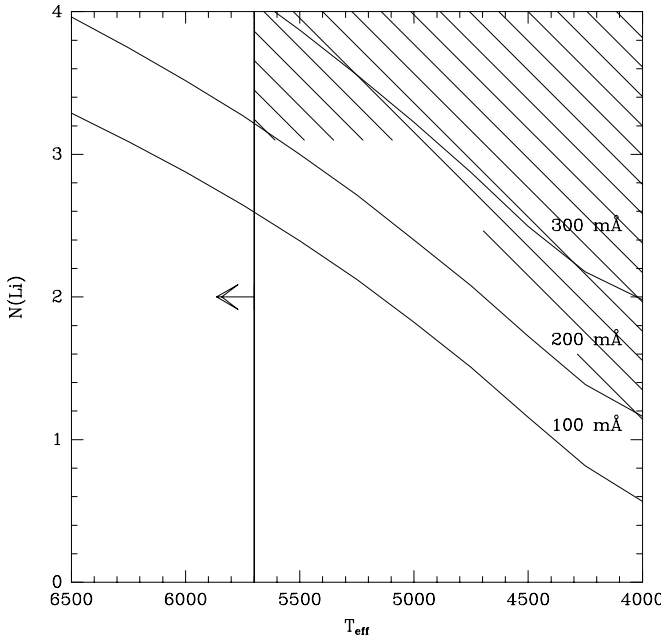, width=8.8cm, bbllx=0pt, bblly=0pt,
      bburx=560pt, bbury=550pt, clip=}
  \end{center}
  \caption{The lithium abundance is shown, as a function of effective
    temperature, for equivalent width values of the \lii\ doublet of
    100, 200 and 300\,m\wl, following the curves of growth of
    Soderblom et al. (1993). The heavy vertical line represents the
    cool \teff\ bound for $\simeq$ G5 stars, and the shaded region is
    the region which is expected to be occupied by bona fide PMS
    stars.}
  \label{fig:cog}
\end{figure}

In Fig.~\ref{fig:cog}, in which we have plotted the lithium abundance
implied by different (true) observed equivalent widths of the \lii\ 
doublet, for an equivalent width of 100, 200 and 300\,m\AA. The region
to the left of the thick vertical line in Fig.~\ref{fig:cog} is the
region occupied by the early- to mid-G dwarfs, for which lithium
cannot be used to separate PMS and main-sequence objects . The
boundary of the dashed region is the approximate upper boundary of the
lithium abundances measured in the Pleiades by \cite*{sjb+93}, so that
any star lying outside the dashed region has lithium abundances
compatible with its being a main sequence star. The region in which
PMS stars can be discriminated from the lithium abundance is thus the
dashed area in Fig.~\ref{fig:cog}, although, given the spread of
lithium abundance observed at any given age, the actual boundary has
to be assumed to be quite fuzzy. Taking a mass-independent threshold
of 100\,m\AA\ for the equivalent width of the \lii\ doublet will
select (as already remarked by \cite{bhs+97}), in addition to whatever
true WTTS there may be in the sample, many stars which are simply
young main sequence stars, thus inflating the detected number of WTTS
with several spurious sources. A threshold of $\simeq 250$\,m\AA\ 
would approximately follow the sloping boundary of the shaded region,
but at the same time it would miss many true WTTS among the hotter
stars, as well as a few at the cooler end.

\section{Where is the young main-sequence population?}
\label{sec:young}

As remarked in the Sect.~\ref{sec:intro}, the RASS-WTTS papers show a
severe lack of normal, young active main-sequence stars in their
samples.  For example, \cite*{aks+95} report (see their Table~4) to
have investigate 112 X-ray sources (in an area of $\simeq 150$ square
deg) around the Chameleon SFR, and, among the 112 X-ray sources
studied, to have found 75 new WTTS and only 10 active stars which are
not classified as WTTS on the basis of their low-resolution spectra.
It is suggestive, in the light of the results of Sect.~\ref{sec:anal}
and~\ref{sec:m}, that the non-WTTS active stars reported by
\cite*{aks+95} are all dMe stars, as it is at these spectral types
that low-resolution spectra have some diagnostic value.  In addition
to this, they report another 13 non-WTT stellar sources which were
previously known, although from their Table~6 it appears that some are
early-type stars and thus non-coronal sources. Thus, 23 non-WTT stars,
or 18, subtracting the early-type stars, in $\simeq 150$ square deg.
For the same sky area \cite*{akc+97} report a limiting sensitivity of
$\log f_{\rm X} = -12.8$ (the same limiting sensitivity can be derived
from the RASS exposure time in the region of $\simeq 2500$\,s reported
by \cite{aks+95}, leading to a limiting PSPC count rate of $\simeq
0.005$\,cts\,s$^{-1}$). A computation based on the model of
\cite*{fms+92} and \cite*{sfm95} predicts at this limiting flux level,
between 60 and 80 active, non-PMS stars, in that area of the sky,
depending on the assumed value for the space density of RS~CVn
binaries. This is a factor of $\simeq 3$ higher than the 23 active
non-PMS stars reported by \cite*{aks+95}, showing that many of the
putative WTTS in their sample are likely to be misclassified
main-sequence stars. Note that the computation discussed here is in
full agreement with the observed numbers in the EMSS, i.e. these large
numbers of young main-sequence objects are not only expected, at the
X-ray flux levels, but their presence has already been verified on
fully identified samples. The \cite*{ghm+96} model, predicts an even
larger number of main-sequence stellar sources, i.e. about one per
square degree at these fluxes and latitudes, approximately two thirds
of which are expected to be of age 1\,Gyr or older (and with a
significant fraction of the ones younger than 1\,Gyr being on the main
sequence). A similar prediction is made by \cite*{bhs+97}. The
\cite*{ghm+96} model has been shown by \cite*{mgh+97} to provide a
good match to the RASS population of the Galactic plane, where they
report 77 coronal sources in a sky area about half the size and at a
limiting flux approximately twice as shallow of the one surveyed by
\cite*{aks+95} --- although at slightly lower galactic latitude.

Later works in the RASS-WTTS line find significantly higher fractions
of non-PMS coronal sources. For example, \cite*{mms+97} find, in the
Tau-Aur region (where the RASS exposure time is $\simeq 500$\,s),
significant numbers of non-PMS coronal sources, even in a sample which
had been optimized for searching for PMS sources. The difference in
methodology is at least in part likely to be the cause of the
difference in the detected source population, and in particular the
larger spectral resolution employed ($\simeq 1$\,\AA), which allows
better discrimination of true WTTS from main-sequence stars.

\section{Discussion}
\label{sec:disc}

We have shown, by analysis of the low-resolution spectra of a number
of low-mass stars spanning a wide range of spectral types as well as
of lithium abundances, and by direct comparison of their low- and
high-resolution spectra, that usage of low-resolution spectra alone is
likely to lead, at least in G and K stars, to high inferred lithium
abundances in late-type stars. An absorption feature at 6708\,\AA\ 
appears to be commonly present, in the low-resolution spectra of G and
K stars, independently from their actual lithium abundance. Such
feature appears to be present, for stars later than $\simeq$ M0, only
in true high-lithium sources. Thus, classification of stellar
counterparts to soft X-ray sources done exclusively on the basis of
low-resolution optical spectroscopy is likely to significantly
over-estimate the number of PMS stars present in the source
population. We have also shown that, even with fully reliable \lii\ 
equivalent widths, the adoption of a single equivalent width threshold
will lead to over-estimate the number of WTTS sources present in the
sample.

As discussed in Sect.~\ref{sec:intro}, several works have recently
appeared in the literature which present the identification of stellar
counterparts to soft X-ray sources based on low-resolution spectra
alone. These works discuss X-ray sources in the general direction of
star forming regions, but usually cover large region of the sky,
extending to quite large projected distances from the SFR. A common
feature to all these works is that they seem to find, in addition to
the expected concentration of PMS stars in and around the SFR, a large
number of widespread WTTS with no apparent immediate relationship with
the SFR under investigation, which, as discussed in
Section~\ref{sec:intro}, are a challenge to current ideas of low-mass
star formation. At the same time, the same samples lack the large
number of young main-sequence coronal sources which are known to be
present in X-ray selected samples at these flux levels.

We make the hypothesis that a non-negligible fraction of the ``field
WTTS'' discussed in the RASS-WTTS papers are normal, active young
low-mass stars, on, or very near to the main sequence.
The arbitrary placement of foreground active stars at the distance of
the putative parent SFR (which is common practice in the WTTS-RASS
papers) will make them appear brighter then they actually are, and
thus make them wrongly appear as still in a PMS contraction phase when
placed on evolutionary tracks.  The apparent large number of dispersed
WTTS are thus most likely not the solution to the still standing
puzzle of the apparent lack of the deficiency of stars older than
$\simeq 2$\,Myr in most known star forming regions (the so-called
missing post-T Tauri problem), as discussed by \cite*{fei96}.
\cite*{pg97} have recently argued that the post-T Tauri problem is a
false one, as it is based on the assumption of a constant
star-formation rate in giant molecular clouds, an assumption which,
based on the similarity between the molecular cloud lifetime and the
ambipolar diffusion time they show to be unlikely. Rather, they argue,
star formation accelerates sharply toward the end of a cloud's
lifetime, thus justifying the lack of large numbers of older PMS in
SFRs.

\cite*{mfs97} have recently used the Hipparcos parallaxes of the
subsample of EMSS and ESS stars which have been observed by Hipparcos
to accurately position these stars in an HR diagram, showing that only
one of the stars in the sample is far away from the main sequence and
clearly still in a contracting phase. The rest of the population is
mostly composed of main-sequence objects, with $\simeq 20$\% giants.
While this subsample suffers from a bias toward brighter stars, and it
is thus lacking many of the fainter and more active stars (some of
which are known to be PMS stars from their lithium abundance), they
show that all of the stars in their sample (seven) which would have
been classified as WTTS using the \cite*{wks+96} 100\,m\AA\ criterium
(even using high-resolution spectra, and thus reliable \lii\ doublet
equivalent widths) are very close to or on the main-sequence. Thus,
while there certainly are a number of bona fide WTT stars in the
sample of active stars selected from the \E\ surveys (as for example
the low-gravity, very high lithium abundance stars of \cite{mmf+96}),
the majority of the low-mass stars appear however to be already on or
very close to the main sequence stage.  Given the similar limiting
sensitivity of the RASS and of the EMSS, the detected source
population has to be similar, and therefore, again, a large fraction
of the RASS stellar sources are expected not to be in the PMS stage
but rather young main-sequence stars.

Obviously, a (perhaps considerable) fraction of the stars identified
in the RASS identification programs discussed above will be true WTTS,
still contracting toward the main sequence, given also the vicinity of
the surveyed areas to SFR's. However, lacking accurate,
high-resolution based lithium abundances and distance measurements,
they cannot be separated from the normal active main-sequence stars
present in the sample. Given the type of biases and their dependence
on the stellar mass, it is likely that the fraction of bona fide WTTS
will be higher among late K and M stars, and lower for F and G stars.
Any definitive assessment of the true nature of these RASS sources and
of the (statistical) properties of RASS-selected PMS populations will
thus have to wait for the availability of high-resolution
spectroscopic data, including measurements of the \lii\ doublet, which
will help in screening the bona fide WTTS sources in the sample, at
least for the cooler stars.

\acknowledgements{G. M. and S. S. acknowledge financial support from
  GNA-CNR, and MURST (Ministero della Universit\`a e della Ricerca
  Scientifica e Tecnologica).  We have extensively used the Simbad
  database in the course of the present work, and the IRAF software
  system for the data reduction.  We would like to thank E. Feigelson
  for some very useful discussions on the subject matter of this
  paper, as well as R. Pallavicini for some useful comments on an
  early draft of this paper. We also thank the referee, Dr. L.\,W.
  Hartmann, for the useful comments which have helped to improve the
  paper.}


\begin{thebibliography}{}

\bibitem[\protect\astroncite{Alcal\'a et~al.}{1995}]{aks+95}
Alcal\'a J.\,M., Krautter J., Schmitt J.\,H.\,M.\,M. et~al. 1995,
A\&AS, 114, 109 

\bibitem[\protect\astroncite{Alcal\'a et~al.}{1996}]{atw+96}
Alcal\'a J.\,M., Terranegra L., Wichmann R. et~al. 1996, A\&AS, 119, 7

\bibitem[\protect\astroncite{Alcal\'a et~al.}{1997}]{akc+97}
Alcal\'a J.\,M., Krautter J., Covino E. et~al. 1997, A\&A, 319, 184

\bibitem[\protect\astroncite{Brice\~no et~al.}{1997}]{bhs+97}
Brice\~no C., Hartmann L.\,W., Stauffer J.\,R., Gagn\'e M., Stern
R.\,A. 1997, AJ, 113, 740

\bibitem[\protect\astroncite{Favata et~al.}{1992}]{fms+92}
Favata F., Micela G., Sciortino S., Vaiana G.~S. 1992, A\&A, 256, 86

\bibitem[\protect\astroncite{Favata et~al.}{1993}]{fbm+93}
Favata F., Barbera M., Micela G., Sciortino S. 1993, A\&A, 277, 428

\bibitem[\protect\astroncite{Favata et~al.}{1995}]{fbm+95}
Favata F., Barbera M., Micela G., Sciortino S. 1995, A\&A, 295, 147

\bibitem[\protect\astroncite{Favata et~al.}{1996}]{fms96a}
Favata F., Micela G., Sciortino S. 1996, A\&A, 311, 951

\bibitem[\protect\astroncite{Feigelson}{1996}]{fei96}
Feigelson E.\,D. 1996, ApJ, 468, 306

\bibitem[\protect\astroncite{Guillot et~al.}{1996}]{ghm+96}
Guillot P., Haywood M., Motch C., Robin A.\,C. 1996, A\&A, 316, 89

\bibitem[\protect\astroncite{Krautter et~al.}{1997}]{kws+97}
Krautter J., Wichmann R., Schmitt J.\,H.\,M.\,M. et~al. 1997, A\&A, in press

\bibitem[\protect\astroncite{Magazz\`u et~al.}{1997}]{mms+97}
Magazz\`u A., Mart\'{\i}n E.\,L., Sterzik M.\,F. et~al. 1997, A\&AS, in press

\bibitem[\protect\astroncite{Micela et~al.}{1993}]{msf93}
Micela G., Sciortino S., Favata F. 1993, ApJ, 412, 618

\bibitem[\protect\astroncite{Micela et~al.}{1995}]{mfp+95}
Micela G., Favata F., Pye J.\,P., Sciortino S. 1995, A\&A, 298, 505

\bibitem[\protect\astroncite{Micela et~al.}{1997}]{mfs97}
Micela G., Favata F., Sciortino S. 1997, A\&A, in press

\bibitem[\protect\astroncite{Morale et~al.}{1996}]{mmf+96}
Morale F., Micela G., Favata F., Sciortino S. 1996, A\&AS, 119, 403

\bibitem[\protect\astroncite{Motch et~al.}{1997}]{mgh+97}
Motch C., Guillot P., Haberl F. et~al. 1997, A\&A, 318, 111

\bibitem[\protect\astroncite{Palla \& Galli}{1997}]{pg97}
Palla F., Galli D. 1997, ApJ, 476, L35

\bibitem[\protect\astroncite{Pasquini et~al.}{1994}]{plp94}
Pasquini L., Liu Q., Pallavicini R. 1994, A\&A, 287, 191

\bibitem[\protect\astroncite{Sciortino et~al.}{1995}]{sfm95}
Sciortino S., Favata F., Micela G. 1995, A\&A, 296, 370

\bibitem[\protect\astroncite{Soderblom et~al.}{1993}]{sjb+93}
Soderblom D.\,R., Jones B.\,F., Balachandran S. et~al. 1993, AJ, 106, 1059

\bibitem[\protect\astroncite{Stauffer et~al.}{1993}]{spg+93}
Stauffer J.\,R., Prosser C.\,F., Giampapa M.\,S., et~al. 1993, AJ, 106, 229

\bibitem[\protect\astroncite{Walter et~al.}{1994}]{wvm+94}
Walter F.\,M., Vrba F.\,J., Mathieu R.\,D., Brown A., Myers
P.\,C. 1994, AJ, 107,  692 

\bibitem[\protect\astroncite{Wichmann et~al.}{1996}]{wks+96}
Wichmann R., Krautter J., Schmitt J.\,H.\,M.\,M. et~al. 1996, A\&A, 312, 439

\bibitem[\protect\astroncite{Wichmann et~al.}{1997}]{wkc+97}
Wichmann R., Krautter J., Covino E. et~al. 1997, A\&A, 320, 185

\end{thebibliography}

\end{document}